\documentclass{emulateapj}

\shorttitle{Magnetic Core-Envelope Explosions}
\shortauthors{Matt, Frank, \& Blackman}

\begin{document}

\title{Astrophysical Explosions Driven by a Rotating, Magnetized,
Gravitating Sphere}

\author{
Sean Matt\altaffilmark{1},
Adam Frank\altaffilmark{2}, and
Eric G.\ Blackman\altaffilmark{2}}




\altaffiltext{1}{Dept. of Astronomy, U. of Virginia, 
Charlottesville VA, 22904; Levinson/VITA Fellow; seanmatt@virginia.edu}

\altaffiltext{2}{Physics \& Astronomy Dept., U. of Rochester, 
Rochester NY, 14627; afrank@pas.rochester.edu, blackman@pas.rochester.edu}

\begin{abstract}

We present the results of a numerical magnetohydrodynamic simulation
that demonstrates a mechanism by which magnetic fields tap rotational
energy of a stellar core and expel the envelope.  Our numerical setup,
designed to focus on the basic physics of the outflow mechanism,
consists of a solid, gravitating sphere, which may represent the
compact core of a star, surrounded by an initially hydrostatic
envelope of ionized gas.  The core is threaded by a dipolar magnetic
field that also permeates the envelope.  At the start of the
simulation, the core begins to rotate at 10\% of the escape speed.
The magnetic field is sufficiently strong to drive a
magneto-rotational explosion, whereby the entire envelope is expelled,
confirming the expectation of analytical models.  Furthermore, the
dipolar nature of the field results in an explosion that is enhanced
simultaneously along the rotation axis (a jet) and along the magnetic
equator.  While the initial condition is simplified, the simulation
approximates circumstances that may arise in astrophysical objects
such as Type II supernovae, gamma ray bursts, and proto-planetary
nebulae.


\end{abstract}


\section{Introduction \label{sec_intro}}

Over the last two decades magnetic fields have been identified as the
principle, universal agent for creating collimated astrophysical
outflows. When a magnetic field is anchored in a rapidly rotating
object near the bottom of a gravitational potential, the field can act
as a drive-belt, tapping rotational kinetic energy and launching
plasma back up the potential well.  This magneto-rotational (MR)
scenario for jet launching has been explored by numerous authors, both
analytically \citep{blandfordpayne82, pelletierpudritz92} and
numerically \citep*{ouyed3ea97, krasnopolsky3ea99}.  These authors
showed this mechanism can produce steady flows of matter, in a sense
that the outflow engine still operates while the flow is
observable---as in jets from young stellar objects
\citep{reipurthbally01} and active galactic nuclei
\citep[][]{begelman3ea84}.

On the other hand, this mechanism can also operate in a transient
event, linked to a rapid evolution of the source and driving an
explosion.  In this case, the engine loses a significant fraction of
its power by the time the outflow (or its interaction with the
environment) is detected.  The proposed mechanism for gamma ray bursts
\citep[GRBs;][]{piran05} and supernovae (SNe), lie in the explosive
regime.  There is also growing evidence \citep{bujarrabalea01} that
collimated outflows from planetary nebulae (PNs) are explosive.

Transient MR explosions have not been explored in as much detail as
the steady-state models.  In a collapsing star, differential rotation
near the core may amplify the field linearly \citep[when turbulence is
unimportant;][]{kluzniakruderman98, wheeler3ea02} or exponentially
\citep[when turbulence is important;][]{akiyamaea03, blackman3ea06}.
When differential rotation twists a poloidal magnetic field,
generating toroidal field $B_\phi$, enough rotational energy may be
tapped to power a supernova explosion.  The role of toroidal field
pressure in then driving an outflow has been highlighted by
\citet{lyndenbell96}, who explored magneto-static ``magnetic towers''
to understand jet properties \citep[see also][]{liea01}, and
\citet{uzdenskymacfadyen06} has extended this work to exploding stars.
Numerical simulations by \citet{leblancwilson70} and more recently by
\citet*{ardeljan3ea05} support the idea that MR explosions can be
important for driving SNe and GRBs.

However, the inherent time-dependence and general complexity of the
system presents a significant challenge to our understanding of the
basic mechanism and identification of key parameters.  It is clear
that a MR explosion is ultimately driven by the rotational kinetic
energy extracted from the material that is left behind (the stellar
remnant).  But, notably, there is still some uncertainty as to whether
an accretion disk must form inside the star \citep[as
in][]{uzdenskymacfadyen06} or whether the rotation of the stellar core
\citep[e.g., a protopulsar;][]{wheeler3ea02} alone will drive a MR
explosion.

In this Letter, we present a magnetohydrodynamic (MHD) simulation of a
MR explosion.  While our setup is simple, similar to the analytical
``protopulsar jet'' model described by \citet{wheeler3ea02}, our
simulation captures the nonlinear dynamics, as the magnetic field is
twisted at the shear layer between the core and envelope.  This
heuristic approach enables a better understanding of the basic MR
physics and complements the more phenomenological approach of
previous, more complex, numerical studies.



\section{Physical Experiment and Numerical Method} \label{sec_setup}

	\subsection{Physics \label{sub_basic}}

The model consists of a gravitating, magnetized sphere (the ``core,''
representing a stellar core) surrounded by an overlying envelope.  The
core is solid in a sense that no mass flows in or out of the surface,
and it rotates as a solid body.  The core is also a conductor with a
rotation-axis-aligned dipole magnetic field anchored to the surface.
This magnetic field couples to the envelope, whose behavior is
characterized by the ideal MHD equations.  This core-envelope model
approximates the conditions in the interior of a star where there is a
very steep density gradient, for example in the proximity of a nearly
degenerate core in an evolved star.  There is an implicit assumption
that the envelope self-gravity is negligible.

We assume an initial state in which the envelope is not rotating and
rests in hydrostatic equilibrium, and whose density falls of as
$r^{-2}$.  The initial dipole magnetic field is force-free and
permeates the envelope.  At a time defined as $t$ = 0 in the
simulations, the core begins rotating at a constant rate of 10\% of
the escape speed.  This initiates a shear at the boundary between the
surface of the core and the base of the envelope.

This strong initial shear and purely poloidal magnetic field
approximates the conditions in a real star immediately following an
abrupt change in structure, such as a core collapse (or envelope
expansion).  If the collapse is rapid enough, $B_\phi$ subsequently
amplified at the core-envelope interface will be much stronger than
$B_\phi$ generated elsewhere before the collapse, and the envelope
rotation will be negligible.  The envelope is unlikely to be at rest
following such a change, and any bulk motion could quantitatively (but
not qualitatively) change our results.  For example, the ram pressure
of an infalling envelope would change the magnetic energy required for
an explosion by approximately a factor of two.  Finally, the field
geometry is reasonable, since pulsars and magnetic white dwarfs are
thought to have dipolar fields (though tilted with respect to their
rotation axes).

The poloidal magnetic field connecting the core to the envelope is
twisted by the shear, generating an azimuthal component to the
magnetic field, $B_\phi$, near the interface, which increases
approximately linearly in time.  The magnetic pressure force
associated with $-\nabla B_\phi^2$ will be directed generally outward
from the core, so when $B_\phi$ becomes dynamically important, the
field will expand, driving the ionized envelope material in front of
it.  If the expansion is powerful enough, it can be explosive and
drive off the entire envelope.  We intentionally inhibit any sort of
wind from the surface of the core, so the outflow in this explosion is
composed of material that originally overlies the core (i.e., envelope
material).  The flow thus depends on the initial conditions existing
above the surface of the core and is inherently a transient
phenomenon.  This contrasts with any steady-state wind theory whose
solution depends only on the boundary conditions at the wind driving
source.

     \subsection{Parameters \label{sub_parms}}

The key parameters in the system are the initial magnetic, rotational
kinetic, gravitational potential, and thermal energy densities near
the core-envelope interface.  It is instructive to compare the
characteristic speeds associated with these energies, namely the
Alfv\'en speed [$v_{\rm A} = B / (4 \pi \rho)^{1/2}$, where $B$ is the
dipole field strength at the equator of the core and $\rho$ is the
mass density at the base of the envelope], rotation speed at the
equator ($v_{\rm rot}$), escape speed at the core surface ($v_{\rm
esc}$), and sound speed at the base of the envelope ($c_{\rm s}$),
respectively.

The assumption of initial hydrostatic equilibrium couples the thermal
energy to the potential energy, effectively reducing the number of
free parameters (by requiring the initial $c_{\rm s} \approx 0.53
v_{\rm esc}$).  Also, since the simulations can be scaled to any
system with similar ratios of energies, there are only two fundamental
parameters, which can be cast as the dimensionless velocity ratios
$v_{\rm A} / v_{\rm esc}$ and $v_{\rm rot} / v_{\rm esc}$.  Here, we
present results from a simulation with $v_{\rm A} / v_{\rm esc} = 1.0$
and $v_{\rm rot} / v_{\rm esc} = 0.1$.

The key parameter that determines whether or not the envelope will be
driven off is $\chi \equiv (v_{\rm A} v_{\rm rot})^{1/2} / v_{\rm
esc}$.  We have run a limited parameter study thus far (not presented
in this Letter), which suggests the threshold for explosion occurs
near $\chi \ga 0.2$.  The model presented here has $\chi \approx 0.3$.

\subsection{Numerics}

We employ the 2.5-dimensional MHD code of \citet{matt02}, and the
reader can find details of the code there and also in \citet{mattea02}
and \citet{mattbalick04}.  The code, solves the ideal (non-resistive)
MHD equations using a two-step Lax-Wendroff, finite difference scheme
\citep{richtmyermorton67} in cylindrical $(\varpi, \phi, z)$ geometry.
The formulation of the equations allows for a polytropic equation of
state (we adopt $\gamma = 5/3$), includes a source term in the
momentum and energy equations for point-source gravity, and assumes
axisymmetry ($\partial / \partial \phi$ = 0 for all quantities).

The computational domain consists of five nested grids (or ``boxes'')
in the cylindrical $\varpi$-$z$ plane.  Each box contains $401 \times
400$ gridpoints (in $\varpi$ and $z$, respectively) with constant grid
spacing.  The boxes are nested concentrically, so that the inner
(first) box represents the smallest domain at the highest resolution.
The next outer box represents twice the domain size with half the
spatial resolution (an so on for other boxes).  The simulation ended
before the explosion propagated beyond the fourth box.  A circular
boundary with a radius of 30.5 gridpoints and centered on the origin
represents the surface of the core.  Thus, the innermost box has a
spatial resolution of 0.033 core radii, $R_{\rm c}$, and a domain size
of 13.1 $R_{\rm c}$, and the fourth box has a resolution and domain
size of 0.26 $R_{\rm c}$ and 105 $R_{\rm c}$, respectively.

\begin{figure}
\epsscale{1.15}
\plotone{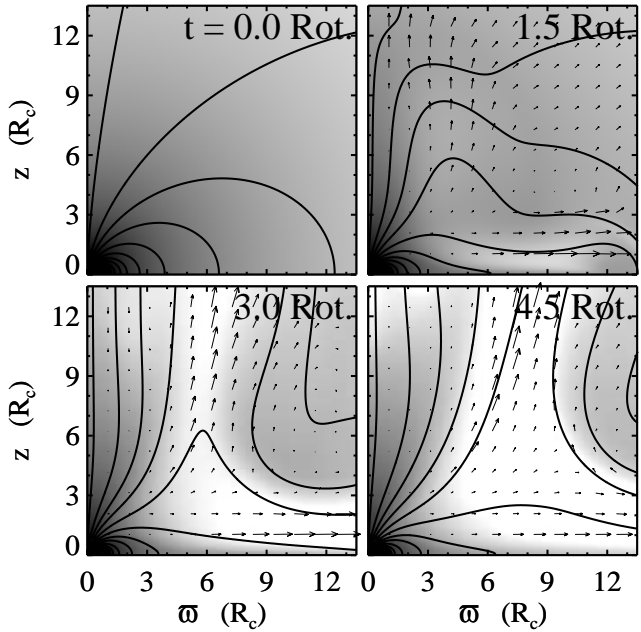}

\caption{Greyscale images of log density (black is highest density),
poloidal field lines, and velocity vectors show the evolution of the
system in the region near the core.  The data are from the first
simulation grid.  The core is in the lower left of each panel, and the
rotation axis is vertical.  The four panels represent a time-sequence
from 0, 1.5, 3.0, and 4.5 rotations of the core.
\label{fig_baseseq_close}}

\end{figure}

We use standard outflow conditions on the outer box boundaries,
appropriate for wind studies.  We also use standard boundary
conditions on the rotation axis ($\varpi = 0$) and reflection symmetry
on the equator ($z =0$).  The simulation domain, then, consists of a
$\varpi$-$z$ slice through a single quadrant.  The surface of the core
is represented by a circular inner boundary, centered at the origin
($\varpi = z = 0$).  Here, we enforce boundary conditions on a three
gridcell layer, and we have performed several tests to ensure that
this inner boundary behaves as appropriate for the surface of a solid,
rotating, conducting and magnetized sphere.

For the current study, we initialize the entire computational domain
with a spherically symmetric hydrostatic envelope, whose density falls
off as $r^{-2}$ (where $r$ is the spherical radius).  The envelope is
initially stagnant (no rotation or motion) and is threaded everywhere
by a dipole magnetic field that is anchored to the core, as shown in
the upper left panel of Figure \ref{fig_baseseq_close}.

\section{Results: Nature of the Explosion \label{sec_results}}

At the start of the simulations, the dipolar magnetic field begins to
wrap up.  For the chosen parameters, $B_\phi$ becomes dynamically
important after a few tenths of a rotation of the core.  The panels of
Figure \ref{fig_baseseq_close} show the evolution during the first 4.5
rotations of the core.  The expansion of newly generated $B_\phi$ is
evident, as it drives the envelope material outward.  This also
expands the poloidal magnetic field (lines), though some field lines
near the equator remain closed, where material is forced and held in
corotation with the core.  The region near the core is evacuated by
the expansion of the magnetic field, indicative of a transient event,
rather than the beginning of a steady wind.

\begin{figure}
\epsscale{1.15}
\plotone{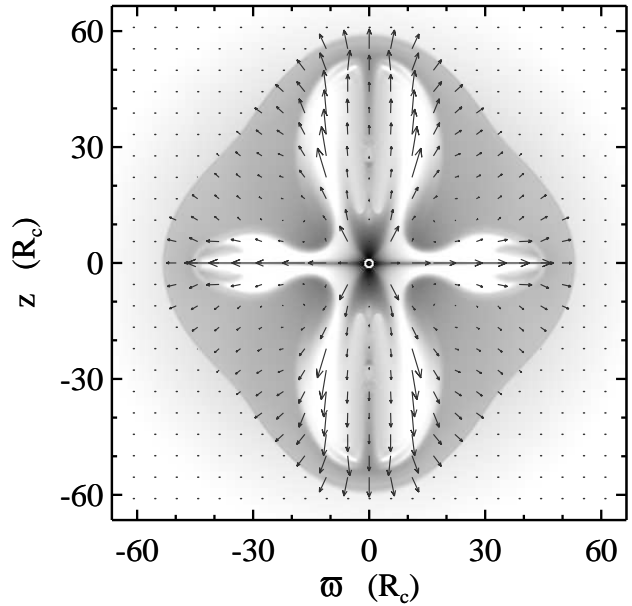}

\caption{Greyscale image of log density (black is highest density) far
from the core, shown after 4.5 rotations of the core.  The data are
from the fourth simulation grid. The core is indicated by a white
circle at the center.  Vectors represent the flow velocity with the
maximum vector length corresponding to 4.1 times $v_{\rm rot}$.
\label{fig_baseseq_far}}

\end{figure}

Figure \ref{fig_baseseq_far} shows the density in the system after 4.5
rotations of the core, on a scale that is 10 times larger than in
Figure \ref{fig_baseseq_close}, and the data have been reflected about
the rotation axis and equator to better illustrate the flow.  In the
Figure, one sees a high-density swept-up shell of envelope material
that is moving outward from the core and bounded by a shock.

The shell has a quadrupolar shape.  This symmetry is due to the
coupling of the rotation to the dipole field, which produces a
$B_\phi$ that has a maximum strength at mid latitudes near the core,
so magnetic pressure forces direct outward from the core, and also
toward the pole and equator \citep[see][]{mattbalick04}.  This shape
can be qualitatively understood as an explosion driven by detonating
two doughnut-shaped charges placed at mid latitudes around the north
and south hemispheres, which results in a convergent blast wave on the
equator and pole.

By 4.5 rotations of the core, the shell material near the rotation
axis and equator is traveling faster than its own internal fast
magnetosonic wave speed, as well as that of the ambient envelope ahead
of the shell.  We expect the expansion speed of the shell to roughly
correlate with $(v_{\rm rot}^2 v_{\rm A})^{1/3}$.  At 4.5 rotations,
the speed of the shell near the rotation axis roughly equals this.
However, the flow speed of the shell material accelerates throughout
the simulation, and does not seem to reach an asymptotic velocity by
the end.  The simulations were stopped after 5.6 rotations of the
core, since the steepening shock front began to produce numerical
instabilities at that time.

The total energy (gravitational potential plus magnetic plus kinetic
plus thermal) integrated over the entire simulation grid (excluding
the core) increases monotonically in time and becomes positive at
around 2.5 rotations.  This means that there is enough energy for the
entire envelope to escape from the gravitational potential well of the
core.  Indeed, by the end of the simulations, the shell material near
the axis and equator is traveling faster than the local escape speed.
On the other hand, material in the shell at mid latitudes is
traveling substantially below its local escape speed.  However, since
this mid latitude material is still accelerating outward at the end of
the simulation, and since the total energy in the system is well above
zero, it appears likely that the entire envelope will escape from the
core (if the simulation were able to run long enough).  This confirms
the general picture of \citet{wheeler3ea02}.  In addition, the
structure of the outflow suggest that, for lower $\chi$, a partial
ejection of the envelope may occur (a ``failed'' explosion) in which
their is not only a jet \citep[as in][]{wheeler3ea02} but also an
equatorial flow, leaving behind the material at mid latitudes.

\begin{figure}
\epsscale{1.15}
\plotone{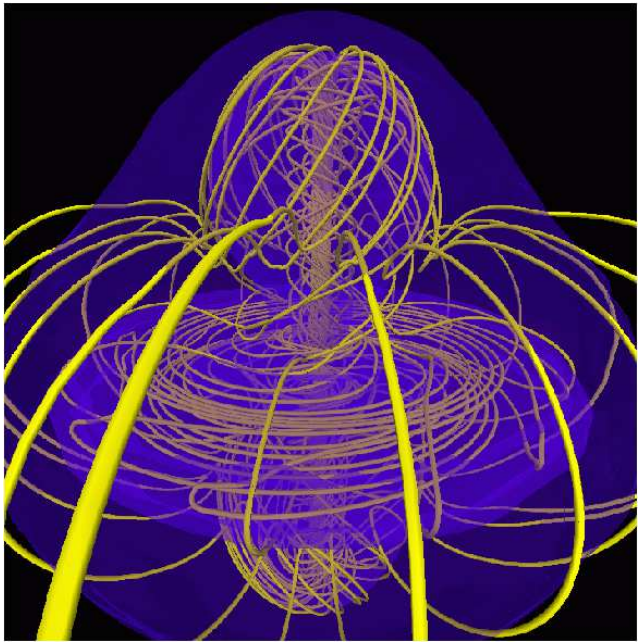}

\caption{Three-dimensional rendering revealing the explosion
mechanism, viewed from $\sim30$ degrees above the magnetic equator.
The two blue surfaces are contours of constant density, each at the
same density value.  The dense shell of swept up envelope material
(see Fig.\ \ref{fig_baseseq_far}) exists between the two surfaces.
Gold wires trace the magnetic field lines and illustrate that the
field is most highly twisted in the low density region interior to the
shell.
\label{fig_3dfar}}

\end{figure}

Finally, the magnetic field configuration after 4.5 rotations is
apparent in Figure \ref{fig_3dfar}, where we have exploited the
axisymmetry of the system to create a 3-dimensional rendering of data
from the fourth simulation grid.  For scale, the outer isodensity
surface corresponds to the outer edge of the shell in Figure
\ref{fig_baseseq_far}.  A comparison between Figures
\ref{fig_baseseq_far} and \ref{fig_3dfar} reveals that the ``hollow''
region interior to the dense shell is actually filled with magnetic
energy, mostly due to the $B_\phi$ component.  The shell is like a
magnetically inflated ``balloon,'' which expands as the spinning core
adds magnetic energy inside.  Thus the spin energy of the core is
ultimately being transferred to the kinetic energy of the envelope
(and the core should spin down, as a result).  The interior region is
magnetically (Poynting flux) dominated, as all material there flows
more slowly than the fast magnetosonic speed.  The flow along the axis
is a magnetic tower jet, similar to what is seen in other simulations
\citep*[e.g.,][]{kato3ea04}, while the flow near the equatorial plane
is more of a ``magnetic pancake,'' a feature of the rotating dipole
field and an interesting new result of this study.

\section{Application to Real Stars \label{sec_discussion}}

Our numerical setup is simple, which allows us to understand the basic
physical principles at work.  The requirements for the mechanism to
operate are a) a shear layer in which the inner region is rotating at
a fraction of breakup speed and b) a magnetic field that results in an
Alfv\'en speed comparable to the escape speed.  Because we focus on
the MR mechanism, and ignore other physics (e.g., neutrino heating),
we can scale our results to various astrophysical systems.  This is
instructive, as it determines what conditions would be necessary for a
MR explosion alone to drive off the envelope.  Here, we briefly apply
our results to the progenitor stars of SNe and PNs.

A simple model of a core-collapse SN suggests that core material
conserves magnetic flux and angular momentum during collapse and
naturally forms a rapidly-rotating, highly-magnetized proto-neutron
star.  We adopt a neutron star (``core'') mass of 1.4 $M_\odot$ and a
core radius of $R_{\rm c} = 10^6$ cm.  If the overlying envelope
contains 1 $M_\odot$ between $r = R_{\rm c}$ and $1 R_\odot$, our
simulation corresponds to a dipole field strength at the core surface
equal to $B = 3 \times 10^{15}$ G.  Note that the envelope mass is
comparable to the core mass, but the self-gravity of the envelope near
the core should not be important, and so the simulation still applies.
In our simulation, we find that the core loses rotational energy at a
constant rate of $B^2 R_{\rm c}^3 / 3 = 3 \times 10^{48}$ ergs per
rotation, which should extract most of the rotational energy of the
core ($\sim 10^{51}$ ergs) in 1 second.  These numbers are consistent
with previous work and should be sufficient to eject the envelope
\citep[e.g.,][]{wheeler3ea02, ardeljan3ea05, blackman3ea06}.

The evolutionary phase preceding PN formation is marked by high mass
loss rates, leading to an expansion of the stellar envelope, which
overlies a proto-white dwarf.  A shear layer at the core-envelope
boundary is expected \citep{blackmanea01}, and as the envelope density
decreases, the Alfv\'en speed should rise.  There may be a threshold
density, below which the remaining envelope may be expelled via a MR
explosion.  We adopt a white dwarf (``core'') mass of 0.5 $M_\odot$,
and $R_{\rm c} = 10^9$ cm.  At this time the envelope is essentially
an extension of the massive outflow, and if it contains 0.05 $M_\odot$
between $r = R_{\rm c}$ and $10^4$ AU, our simulation corresponds to
$B = 9 \times 10^6$ G.  Most of the rotational energy of the core
($\sim 10^{48}$ ergs) should be transferred to the envelope in a few
hundred years.  The linear momentum of the swept up shell will be
$\sim 10^{39}$ g cm s$^{-1}$, which is consistent with the
observations of proto-PNs by \citet{bujarrabalea01}.

Finally, we note that the large-scale field in our simulations is
strong enough to suppress turbulence due to shear instabilities near
the core, and the field therefore is neither subject to decay nor the
exponential growth that is important in some other models.  We have
instead focused on the physics of how the field actually drives the
outflow.


\acknowledgements

This research was supported by NSERC of Canada, McMaster U., and
through a CITA National Fellowship and by the U. of Virginia through a
Levinson/VITA fellowship.



\end{document}